\documentclass[pdflatex,sn-mathphys-num]{sn-jnl}

\usepackage{graphicx}%
\usepackage{multirow}%
\usepackage{amsmath,amssymb,amsfonts}%
\usepackage{amsthm}%
\usepackage{mathrsfs}%
\usepackage[title]{appendix}%
\usepackage{xcolor}%
\usepackage{textcomp}%
\usepackage{manyfoot}%
\usepackage{booktabs}%
\usepackage{algorithm}%
\usepackage{algorithmicx}%
\usepackage{algpseudocode}%
\usepackage{listings}%

\theoremstyle{thmstyleone}%
\theoremstyle{thmstyletwo}%

\theoremstyle{thmstylethree}%

\begin{document}

\title{Systemic Trade Risk Suppresses Comparative Advantage in Rare Earth Dependent Industries}


\author*[1,2,3,4]{\fnm{Peter} \sur{Klimek}}\email{peter.klimek@ascii.ac.at}

\author[1,2]{\fnm{Sophia} \sur{Baum}}

\author[1,5]{\fnm{Markus} \sur{Gerschberger}}

\author[1]{\fnm{Maximillian} \sur{Hess}}

\affil[1]{\orgname{Supply Chain Intelligence Institute Austria}, \orgaddress{\street{Metternichgasse 8}, \city{Vienna}, \postcode{1030}, \country{Austria}}}

\affil[2]{\orgname{Complexity Science Hub}, \orgaddress{\street{Metternichgasse 8}, \city{Vienna}, \postcode{1030}, \country{Austria}}}

\affil[3]{\orgdiv{Institute of the Science of Complex Systems, CeDAS}, \orgname{Medical University of Vienna}, \orgaddress{\street{Spitalgasse 23}, \city{Vienna}, \postcode{1090},\country{Austria}}}

\affil[4]{\orgdiv{Division of Insurance Medicine, Department of Clinical Neuroscience}, \orgname{Karolinska Institutet}, \orgaddress{\city{Stockholm}, \postcode{17177},\country{Sweden}}}

\affil[5]{\orgdiv{Josef Ressel Centre for Real-Time Value Network Visibility, Logistikum},\orgname{University of Applied Sciences Upper Austria}, \orgaddress{\street{Wehrgrabengasse 1-3}, \city{Steyr}, \postcode{4400}, \country{Austria}}}


\abstract{
Rare earth elements (REEs) are critical to a wide range of clean and high-tech applications, yet global trade dependencies expose countries to vulnerabilities across production networks. Here, we construct a multi-tiered input--output trade network spanning 168 REE-related product codes from 2007–2023 using a novel AI-augmented statistical framework. We identify significant differences between dependencies in upstream and intermediate (input) products, revealing that exposure and supplier concentration are systematically higher in input products, while systemic trade risk is lower, suggesting localized vulnerabilities. By computing network-based dependency indicators across countries and over time, we classify economies into five distinct clusters that capture structural differences in rare-earth reliance. China dominates the low-risk, high-influence cluster, while the EU and US remain vulnerable at intermediate tiers. Regression analyses show that high exposure across all products predicts future export strength, consistent with import substitution. However, high systemic trade risk in input products like magnets, advanced ceramics or phosphors, significantly impedes the development of comparative advantage. These results demonstrate that the structure of strategic dependencies is tier-specific, with critical implications for industrial resilience and policy design. Effective mitigation strategies must move beyond raw material access and directly address country-specific chokepoints in midstream processing and critical input production.}

\maketitle

\section{Introduction}\label{sec1}

Rare earth elements enhance efficiency, durability, and miniaturization in a wide range of advanced products. For example, dysprosium, terbium and neodymium and praseodymium alloyed with other metals form powerful permanent magnets used in electric vehicle motors and wind turbines, where they significantly improve performance while reducing weight and size \cite{balaram2019rare, filho2023understanding, dushyantha2020story}. For instance, 95\% of all traction motors\cite{gauss2021rare} contain rare earth magnets and the annual demand will and their critical role spans civilian and military technologies, underscoring their strategic importance.  Demand has grown sharply: global consumption of rare earth oxides rose from 75,500 tonnes (t) in 2000 to 123,100 t in 2016 \cite{goodenough2018rare}. 
Dysprosium and terbium belong to the category of heavy rare earth elements and are therefore of particular importance. These elements are also essential in high-resolution displays, fiber-optic communication systems, and laser guidance modules for defense applications. 

China dominates the rare earth supply chain, accounting for more than 91\% of global refined rare earth output as of 2024 \cite{IEA20205global} and 98\% of Europe's demand is satisfied by China \cite{gauss2021rare}. Although rare earth elements are relatively abundant in the Earth’s crust, the technical and environmental challenges of refining concentrate production in China \cite{tukker2014rare, balaram2019rare}. In 2010, China reduced its export quota by 40\%, capping annual exports at approximately 30,000t—down from nearly 50,000t in the previous year—and briefly halted exports to Japan during a diplomatic dispute over the Senkaku Islands \cite{gavin2013china}. That action triggered global supply concerns and highlighted China’s geopolitical leverage.

Governments responded with several policy interventions to reduce dependencies on Chinese rare earths. In the United States, the Mountain Pass mine in California resumed operations in 2017 and by 2020 supplied around 15\% of global rare earth oxide—although most of the concentrate was sent to China for processing \cite{kramer2021us}. The European Union enacted the Critical Raw Materials Act \cite{ragonnaud2023critical}, launched stockpiles, invested in new primary sources (e.g., Kiruna in Sweden) and supported recycling and substitution research \cite{rizos2022developing}. Nonetheless, China’s dominance persists: even with recent efforts, its share of global refining is projected to remain above 85\% by 2040 \cite{IEA20205global}. Thus, the long-term effectiveness of these interventions remains uncertain.

Research increasingly emphasizes the need to assess rare earth element (REE) criticality through systemic and network-based lenses. Early studies applied a three-dimensional framework combining supply risk, vulnerability to supply restriction, and environmental impact to gauge raw material criticality across metals including REEs \cite{graedel2015materials, graedel2015criticality}. Network metrics on trade flows extended such frameworks to model shock propagation risks through global value chains \cite{klimek2015systemic}. 

More recent studies highlight the complexity of REE flows beyond raw materials \cite{filho2023understanding, sprecher2017novel, alonso2023mapping, schrijvers2020review, mouloudi2022critical}. This includes element-level models of REE flows into selected final-use products such as electric vehicle motors and wind turbines, highlighting the difficulty of tracing REE content through multi-tier supply chains due to data limitations and mixed-oxide reporting conventions \cite{alonso2023mapping}. Other works reassessed criticality of fifteen REEs using a refined three-dimensional framework, finding that praseodymium, neodymium, and dysprosium pose especially high supply risks and vulnerability to supply restriction in electronics and new-energy equipment manufacturing \cite{kosajan2025criticality}.

Telecoupling frameworks place REE trade within socio-ecological context. A recent review identified fragmented themes in prior literature and calling for integrated system-level governance that reflects supply-chain and geographic disconnects across extraction, refining, use, and impact \cite{agusdinata2022critical}. Another critical review likewise found that assessing REE supply security demands integrating circular economy strategies, domestic supply development, and resilience theory to address limitations of isolated risk assessments \cite{salim2022critical}.

Systematic reviews reinforce the methodological challenges in criticality assessments. A survey of 167 studies on critical materials, focusing on gaps in transparency, data quality, and methodological consistency, called for formal inclusion of supply chain risk in material criticality frameworks \cite{mouloudi2022critical}. Similarly, another review emphasized how divergent scopes and unclear indicator logic across criticality methods limit interpretability and comparability of assessments \cite{schrijvers2020review}.

Together, these works underscore that while raw-material-focused evaluations provide foundational insight, they often miss how dependencies propagate through production networks. Ecological and network methods offer richer insight but generally focus on limited products or specific elements. However, our ability to map comprehensive input–output relations across broader product networks at scale has recently significantly advanced with the adoption of novel AI-augmented statistical approaches \cite{fetzer2024ai,karbevska2025mapping}. These advances, however, still fall short of linking rare‑earth exposure to export capacity across technological sectors. 

Here, we address this gap by constructing an AI-augmented, multilayer trade‑and‑production network approach for 168 rare-earth-related product groups over 2007–2023. We quantify exposure, supplier concentration, and systemic trade risk and assess how these factors relate to countries’ development of comparative export strengths in rare-earth-intensive sectors. Our study, therefore, seeks to advance our understanding of how embedded dependencies shape economic resilience and strategic capacity.

\section{Methods}\label{sec2}

We conceptualize dependencies on rare earth elements (REEs) as emerging not only from the extraction of raw materials but also through their integration into processed inputs across complex production chains. To capture these dependencies, we define a multidimensional framework consisting of {\em exposure}, {\em import concentration}, and {\em systemic trade risk}, each reflecting a distinct dimension of vulnerability and interdependence.

We compiled bilateral trade data spanning 2007–2023 from the BACI dataset \cite{gaulier2010baci}, which provides harmonized HS6-level trade flows using the 2007 nomenclature. To enrich this with production context, we mapped HS6 codes representing rare-earth raw materials and downstream products, with seed nodes 280530 (rare-earth oxides), 284610 (cerium compounds) and 284690 (mixed rare-earth compounds), using an AI-augmented, statistical approach, following Baum et al. (2025) \cite{baum2025mapping}. This identifies additional HS6 product codes linked to critical raw materials across extraction, processing, manufacturing, and final-use stages. We computed semantic embeddings of LLM responses describing each HS6 code involved in REE-based products manufacturing, alongside embeddings of official HS6 product descriptions, iterating the process ten times for each material–product combination. Each identified potential link was tested across ten independent prompts to mitigate spurious associations.

For each candidate input–output product pair (i,j), we calculated the Pearson correlation coefficient $\rho_{i,j} = \mathrm{corr}\left(\sum_{a} T_{a,b}^i,\;\sum_{c} T_{b,c}^j\right)
$, where $T_{a,b}^i$ denotes imports of product $i$ from country $a$ to country $b$, and $T_{b,c}^j$ denotes exports of product $j$ from country $b$ to country $c$. To test the null hypothesis $H_0: \rho_{i,j} = 0$, we employed a nonparametric permutation test, holding the input product $i$ fixed and estimating the empirical distribution of $\rho$ values by randomly permuting the output products $j$. We computed a Z-score for each candidate link by comparing its observed $\rho_{i,j}$ against this null distribution. We retained only those links that consistently appeared across multiple LLM-generated embeddings and exhibited statistically significant correlations, thereby reducing the risk of random hallucination. This procedure yielded a directed acyclic production network of 168 HS6 product codes. Each product’s {\em tier} is defined as the shortest path length from a seed node. We weighted each edge in the trade network by recorded bilateral trade value, capturing both volume and directional dependencies.

To quantify REE-related dependencies, we developed three indicators. {\em Exposure} is defined as
$$
\text{Exposure}_{c}^p(t) = \frac{\sum_{a} T_{a,c}^p(t)}{\sum_{a} T_{a,c}^p(t) + \sum_{b} T_{c,b}^p(t)}.
$$
{\em Import concentration} is measured using the Herfindahl–Hirschman Index,
$$
H_{c}^p(t)=\sum_a \left( s_{a,c}^p(t)\right)^2,
$$
with $s_{a,c}^p(t)$ denoting the share of supplier $a$ in country $c$’s import of product $p$ at year $t$. {\em Systemic trade risk} (STR) is defined following a network propagation metric \cite{klimek2015systemic}. It reflects a country’s vulnerability to export disruptions from direct and indirect trading partners, based on trade data and the \textit{Political Stability and Absence of Violence/Terrorism} (PV) indicator provided by the World Bank \cite{worldbank_wgi}:

$$
V_{a,c}^p (t) = \left(1 - \frac{PV(a,t)}{100}\right)\frac{T_{a,c}^p(t)}{\sum_a T_{a,c} ^p(t)} \quad,
$$

$$
STR^p_c(t)=\sum_a (\mathbb{I} - V(t)^p)^{-1}_{c,a}, \quad.
$$

We computed each of these indicators across countries, products, and years. We further reduced them using principal component analysis (PCA) to derive a composite dependency score per country–product–year.

We captured {\em comparative export strengths} via a Revealed Comparative Advantage (RCA) index and calculated a country’s {\em influence} as its marginal contribution to the systemic trade risk experienced by other countries (that is, the contribution of the $a$th summand to  $STR^p_c(t)$). A product was deemed a comparative strength for country $c$ if $\langle \mathrm{RCA}_{c}^p(t)>1 \rangle_t$ holds, where $\langle \cdot \rangle_t$ denotes the average over all years. For each such product, we recorded the dependency indicators for its {\em input products} as defined in the production network. More specifically, input products are identified for each country as those products that have direct input relations with a product in which the country has a comparative advantage.

To identify country typologies, we performed unsupervised clustering on country-year “dependency profiles.” Each profile combined weighted averages of the three dependency indicators (and PCA and influence scores) stratified by network tier (i.e., taking the weighted average over all products with a given tier indicator as derived from the network), with weights proportional to total import and export volumes. Hence, the profiles for each country and year consisted of two sets of exposure, concentration and systemic trade risk indicators (one set for all products, one set for country-specific input products), next to PCA and influence scores for each tier. We first reduced dimensionality using UMAP, then applied density-based clustering (DBM) \cite{meehan2025mudflow}. Each country was assigned to the cluster to which it most frequently belonged across years.

Finally, we tested the relationship between dependency profiles and changes in comparative advantage. We regressed the change in RCA over the 2020–2023 period on dependency profiles calculated over 2008–2011. We conducted a parameter sweep over different observation and projection windows to confirm the robustness of our findings.

\section{Results}\label{sec3}

\subsection{Country-Level Dependencies and Cluster Typology}

\begin{figure}[tbph]
\centering
\includegraphics[width=\textwidth]{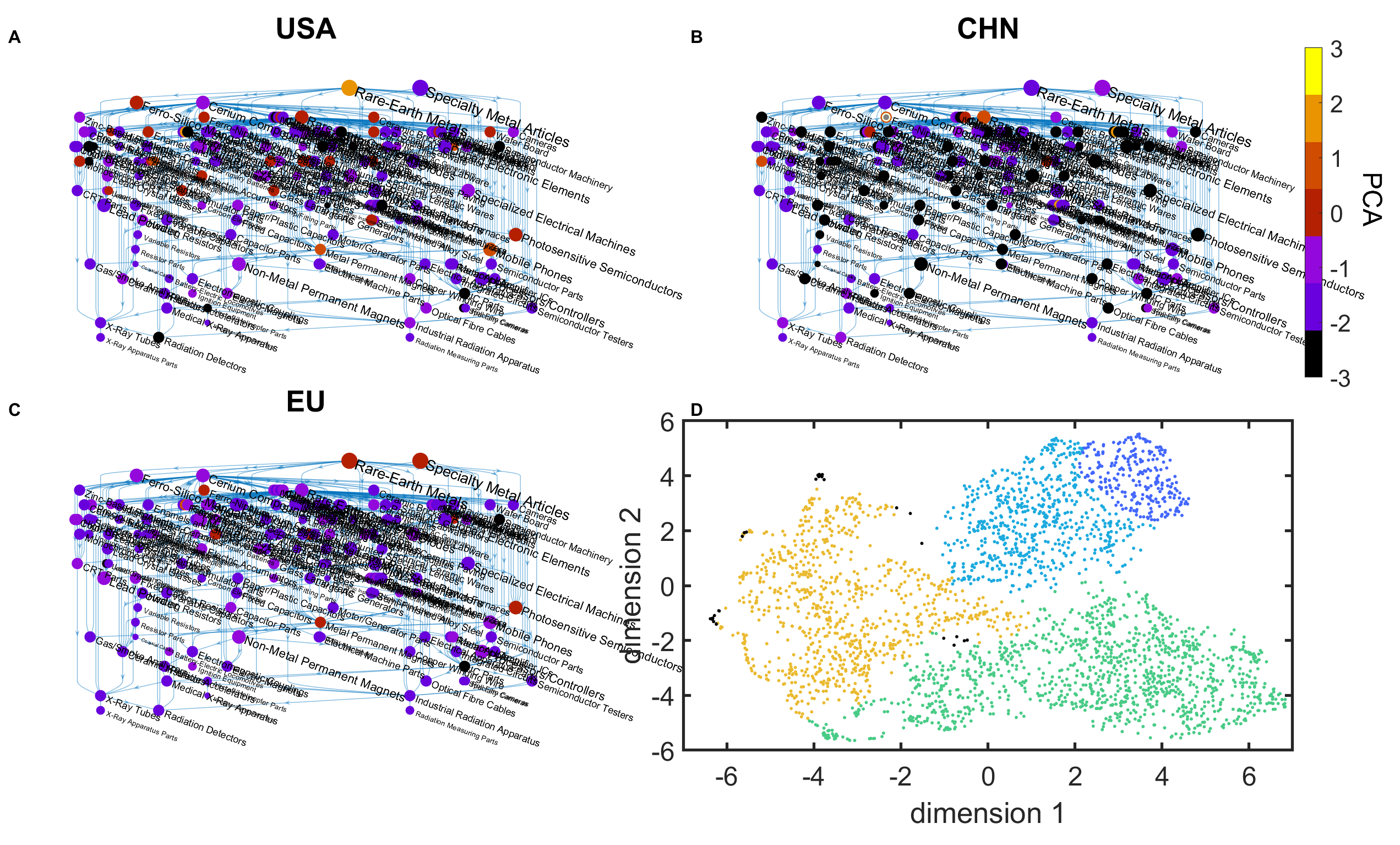}
\caption{Rare-Earth Production Networks and Cluster Typology. Directed acyclic trade networks for rare-earth-related products for the United States (A), China (B), and the European Union (C). Nodes represent HS6 product codes and are organized by production tier from raw materials (Tier 0) to final-use products. Node size is inverse proportional to the product's tier (distance from rare earth nodes in the production network). Colors indicate composite dependency scores derived from principal component analysis (PCA) of exposure, import concentration, and systemic trade risk (black = low dependency; yellow = high dependency). (D) Two-dimensional embedding of country-year observations using UMAP based on dependency profiles. Each point represents one country in one year. Five clusters are identified using DBM. Here colors denote cluster membership. Clusters correspond to distinct structural positions in the rare-earth trade network.}\label{fig1}
\end{figure}

Figure~\ref{fig1} illustrates the rare-earth-related production networks for three major regions: (A) the United States, (B) China, and (C) the European Union. Each network graph is structured hierarchically, beginning with rare earth raw materials at the top and proceeding downward through successive tiers of the value chain. The zeroth tier contains the seed nodes, specifically HS codes 280530, 284610 and 284690, which represent rare earth oxides and compounds. The first tier consists of products that retain significant embedded rare earth content, such as different types of ceramics. Subsequent tiers trace the transformation of these inputs into increasingly complex goods, including permanent magnets, batteries, electronics, specialized machinery, and clean energy technologies.

In each regional network, node size is scaled inverse proportionally to its tier to indirectly reflect embedded rare earth content. This approach ensures a more interpretable visual representation of both trade volume and material significance. Color encodes composite dependency scores derived from a principal component analysis (PCA) of the exposure, import concentration, and systemic trade risk metrics. Colors range from black, indicating low composite dependency, to yellow, signifying high dependency.

The contrast across regions is visually striking. China's network consistently exhibits darker colors, signaling uniformly lower dependency scores throughout the value chain. In contrast, both the United States and the European Union display elevated dependency levels, particularly in the lower tiers of the network. These observations suggest that Western economies experience greater systemic risk and supply vulnerability, especially in raw materials and other lower tier products.

We computed dependency profiles for each country individually. We then constructed an EU dependency profile as a weighted average of all EU member states, with the weights proportional to each country's export volume of a given product.

To formalize these patterns, we conducted an unsupervised cluster analysis of the dependency profiles (one observation corresponds to a country-year combination). The analysis yielded five distinct clusters: four major groupings denoted by color gradients from yellow to blue, and a fifth outlier cluster marked in black. These clusters reflect systematic differences in countries’ dependency profiles across time and tiers. The results of the dimensionality reduction and clustering process are shown in Figure~\ref{fig1}D. Each point represents a country-year observation in the reduced space, with color showing cluster membership.

\subsection{Cluster Dependency Profiles}

\begin{figure}[tbph]
\centering
\includegraphics[width=\textwidth]{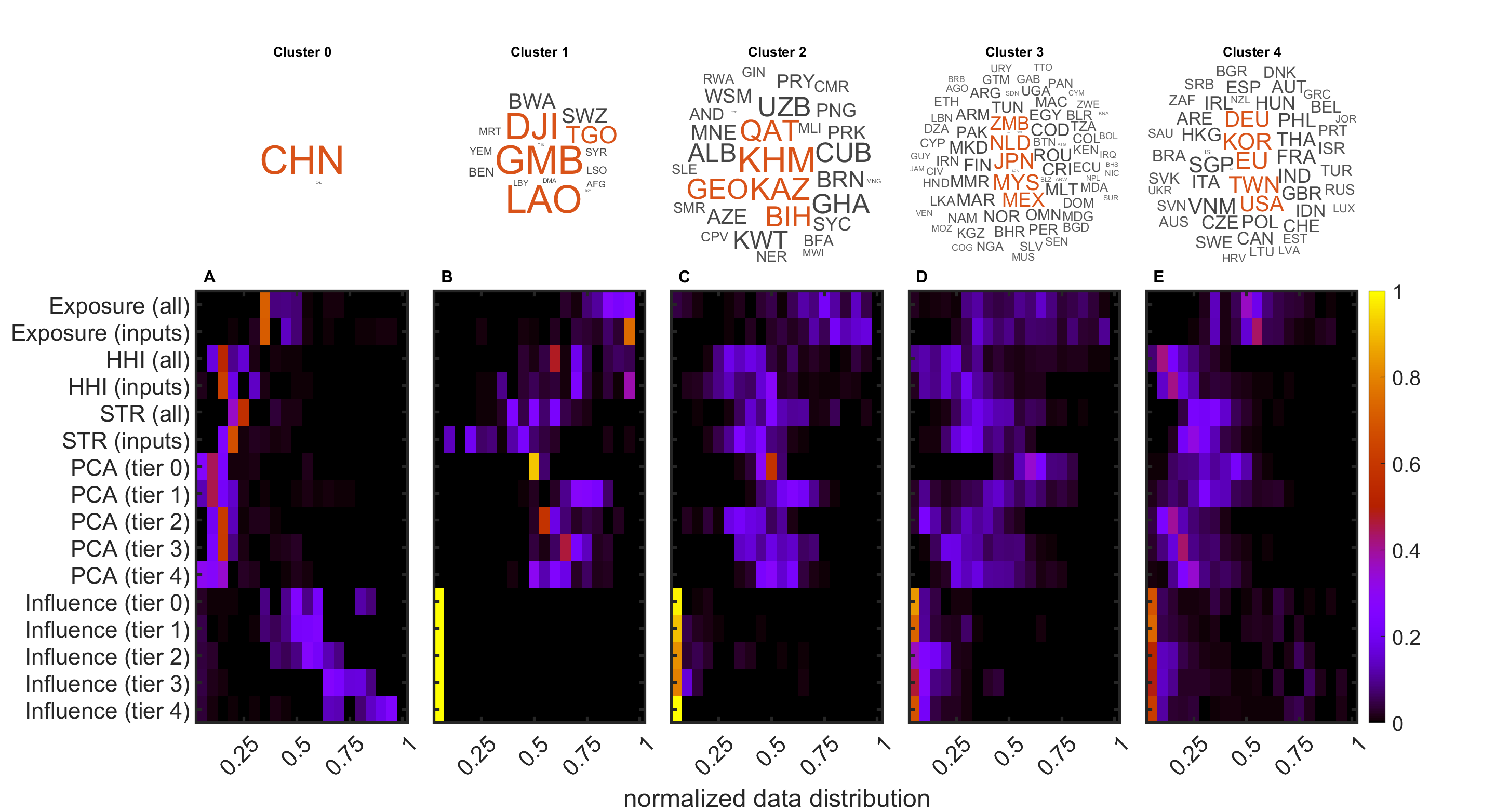}
\caption{Normalized and export-weighted distribution of dependency indicators for each identified cluster. Panels correspond to Clusters 0 to 4. Each histogram displays values for exposure, import concentration, systemic trade risk, influence, and PCA scores across all products and input products, stratified by production tier. Word clouds list the countries in each cluster, with font size proportional to their total export volume across all rare-earth-related products. Cluster 0 (China-dominated) exhibits uniformly low dependency scores and high influence. Clusters 1 and 2 (low-income, high-dependency economies) show high exposure and risk, while Clusters 3 and 4 (advanced and industrialized economies) show decreasing risk levels and growing systemic influence at intermediate tiers.}\label{fig2}
\end{figure}

Figure~\ref{fig2} provides a detailed characterization of each cluster using normalized, export-weighted distributions for all variables included in the clustering process. Each panel represents one cluster, with word clouds indicating the relative export volumes of constituent countries. Larger font size corresponds to greater export value, which also serves as the weighting scheme for the histograms.

Cluster 0 is dominated by China and also includes Chile. It is characterized by uniformly low exposure, low import concentration, and low systemic trade risk across all products and all input product subsets. These values remain stable across all tiers. However, this cluster stands out with significantly higher influence scores, which increase with distance from the raw material tier. This indicates that countries in this cluster exert considerable systemic influence on global supply chains without experiencing corresponding vulnerabilities themselves.

Cluster 1 includes countries such as Gambia, Djibouti, and Laos, alongside other small economies in Africa and Asia. It exhibits the highest values of exposure and concentration, particularly in input products, and elevated systemic trade risk scores. Influence levels are negligible. Notably, dependency scores in this cluster are elevated across all tiers, with slightly lower values at the zeroth tier—likely due to low direct engagement with raw material exports or imports.

Cluster 2 comprises countries like Cambodia, Qatar, Kazakhstan, Georgia, and Bosnia. Its profile closely resembles Cluster 1, though dependency scores are slightly lower and influence marginally higher, reflecting modest integration into global value chains.

Cluster 3 contains more diversified economies such as Japan, Malaysia, Zambia, and Mexico, next to some European country like the Netherlands. This cluster exhibits further reductions in dependency scores, especially with regard to import concentration. While systemic trade risk remains notable, it is increasingly concentrated in the lower tiers of the production network. Influence scores also begin to rise, particularly in intermediate goods.

Cluster 4 includes the European Union, the United States, South Korea, and Taiwan, and reflects advanced industrial economies with relatively sophisticated technological sectors. This cluster continues the trend of declining dependency scores, though these remain substantially higher than those observed in Cluster 0. A bimodal distribution emerges in the influence scores: while countries exhibit low influence in the lowest and highest tiers, intermediate tiers (e.g., processed materials and components) show increased systemic influence.

A consistent theme across all clusters is the asymmetry between input product dependencies and overall dependencies. Systemic trade risk is generally lower for input products than for others, suggesting some resilience at the early stages of the value chain. However, exposure and concentration levels are often higher for these inputs, indicating reliance on few suppliers for strategically important intermediate goods.

\subsection{Trends in Exposure and Import Concentration}

\begin{figure}[tbph]
\centering
\includegraphics[width=\textwidth]{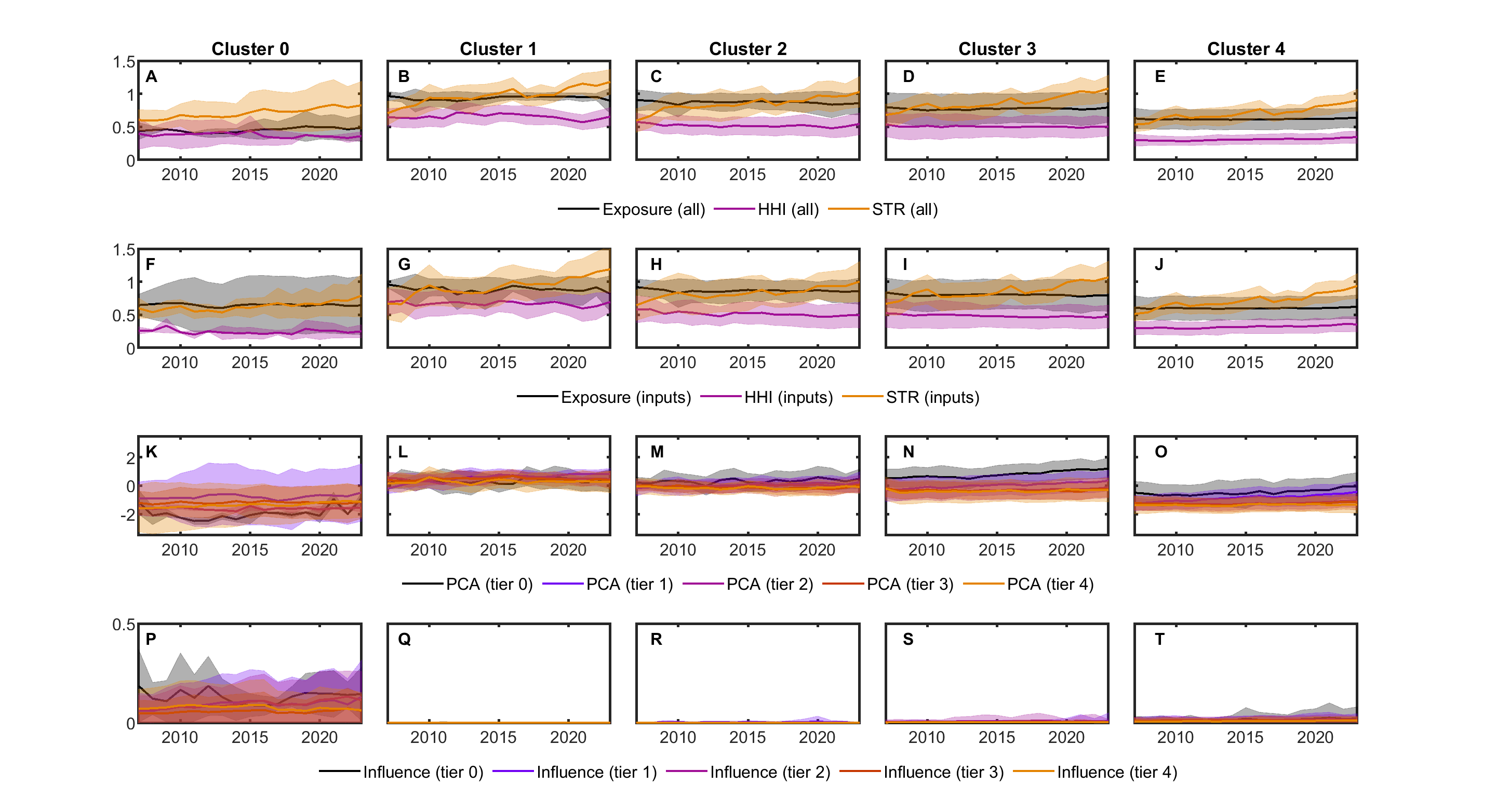}
\caption{Temporal Trends in Dependency Indicators by Cluster. Time series of dependency metrics from 2007 to 2023, stratified by cluster. Each subplot presents the mean and one-standard-deviation confidence interval of the respective dependency indicator (exposure, concentration, systemic trade risk), both for all products and for input products only, next to the PCA score and influence by tier. Exposure and concentration remained relatively stable over time. Systemic trade risk shows a consistent upward trend across all clusters, reflecting increasing reliance on global processing hubs. PCA trends vary by cluster and tier, with raw material tiers becoming more critical in Clusters 3 and 4. Influence scores rise over time in intermediate tiers for advanced economies.}\label{fig3}
\end{figure}

Figure~\ref{fig3} tracks the evolution of the dependency indicators from 2007 to 2023, disaggregated by cluster. Each subplot presents time trends for one group of indicators—exposure, import concentration, and systemic trade risk—for all products and for inputs only. Mean values are shown alongside confidence intervals defined as one standard deviation.

Both exposure levels and import concentration remained relatively stable across time within all clusters. This stability holds whether we consider all products or input-specific dependencies. However, systemic trade risk displays a clear and consistent upward trend across all clusters. This increase reflects growing reliance on a small number of global processing hubs, a trend that is not readily visible when only direct supplier relationships are considered.

The PCA scores show divergent trends by cluster and tier. Clusters 3 and 4—comprising middle- and high-income industrialized nations—show increasing PCA scores at the lower tiers, pointing to rising vulnerabilities in early-stage processing and refining. In the other clusters, there are no clear trends across tiers.

Influence scores also exhibit a moderate upward trend, especially in Clusters 3 and 4. This suggests that advanced economies are gradually exerting more systemic influence in certain parts of the rare-earth-related production network, particularly in lower product tiers.  In Cluster 0 (China), composite influence scores increase slightly at higher tiers, reflecting a shift toward more advanced product engagement.

\subsection{Impact on Comparative Advantage Development}

\begin{table}[tbph]
\caption{Regression Analysis of Comparative Advantage Development. 
Results of fixed-effects regression models estimating the relationship between baseline dependency indicators and changes in revealed comparative advantage (RCA) from 2008–2011 to 2020–2023. Controls include cluster membership and production tier. Significant positive coefficients for exposure in all products suggest a trend toward import substitution. High baseline import concentration and systemic trade risk are negatively associated with RCA development, especially in input products. Influence and existing comparative strength in input products are positively associated with RCA growth. RCA increases with tier and is highest in Cluster 4 (EU, USA, Korea, Taiwan) and lowest in Cluster 1 (low-income, high-risk economies).}\label{tab1}%
\begin{tabular}{p{8cm}lll@{}}
\toprule
Variable & Coefficient  & p-Value \\
\midrule
constant    & -2.8(3)   & $<10^{-4}$   \\
RCA    & -0.13(1)   & $<10^{-4}$   \\
Exposure (all)    & 1.1(1)   & $<10^{-4}$   \\
HHI (all)   & -0.66(10)   & $<10^{-4}$   \\
Systemic Trade Risk (all)    & -0.36(7)   & $<10^{-4}$   \\
Exposure (inputs)    & 0.1(1)   & $0.4$   \\
HHI (inputs)   & 0.26(10)   & $<10^{-4}$   \\
Systemic Trade Risk (inputs)    & -0.37(8)   & $<10^{-4}$   \\
RCA (inputs)    & 0.07(1)   & $<10^{-4}$   \\
Cluster 0    & 1.1(2)   & $<10^{-4}$   \\
Cluster 1    & Ref.   &  \\
Cluster 2    & 0.96(25)   & $0.0001$   \\
Cluster 3    & 1.1(2)   & $<10^{-4}$   \\
Cluster 4    & 1.3(2)   & $<10^{-4}$   \\ Tier 0    & Ref.   &    \\
Tier 1    & 1.1(2)  &  $<10^{-4}$ \\
Tier 2    & 1.1(2)   & $<10^{-4}$    \\
Tier 3    & 1.3(2)   & $<10^{-4}$   \\
Tier 4    & 1.4(2)   & $<10^{-4}$   \\
\botrule
\end{tabular}
\end{table}

Table~\ref{tab1} presents the results of a regression analysis designed to evaluate the relationship between countries' dependency profiles and their ability to develop comparative advantages in rare-earth-related products. The dependent variable is the change in revealed comparative advantage (RCA) between 2008–2011 and 2020–2023. Fixed effects were included for cluster membership and tier level, with Cluster 1 and Tier 0 serving as reference categories.

The analysis shows that countries are more likely to increase their RCA in a product when their baseline RCA is low but exposure is high, suggesting that some nations pursue import substitution strategies that result in strengthened export capabilities. Conversely, high baseline values for concentration and systemic trade risk correlate negatively with RCA growth, indicating that countries embedded in fragile or monopolistic supply chains struggle to enhance their competitiveness.

These relationships differ when focusing on input products. Baseline exposure in input products does not show a significant effect. However, high import concentration and strong existing comparative advantage in input products correlate positively with subsequent RCA growth. In contrast, high systemic trade risk in input products is significantly associated with lower RCA growth, underscoring the critical role that resilient intermediate input networks play in enabling export growth in downstream sectors.

The effect of tier also emerges as important: RCA increases are more pronounced at higher production network tiers. Additionally, countries in Cluster 4—representing advanced economies—exhibit the strongest RCA growth, while those in Cluster 1 show the weakest, reinforcing the idea that embedded systemic risk and limited influence are key barriers to developing rare-earth-related export capacities.

\section{Discussion}\label{sec12}

We offer a novel contribution to the growing literature on critical raw material dependencies by introducing a multilayer, network-based framework that explicitly accounts for indirect, systemic forms of dependency arising from complex production relationships. Rare earth elements (REEs), central to the green energy transition, advanced electronics, and defense applications, are frequently assessed in terms of direct trade volumes or static reserve estimates. However, these conventional approaches neglect how rare earth inputs cascade through global value chains. By capturing these dependencies across tiers of production and time, our findings reveal structural vulnerabilities that are invisible to traditional scalar indicators.

Our analysis reveals how rare earth dependencies are stratified across tiers of the production network. Scalar indicators fail to capture the compounding effects of indirect dependencies. Our tier-based network structure shows that systemic trade risks can be particularly acute not at the level of raw materials (tier 0), but in downstream applications (tiers 2–3), where embedded REE content is still high and substitution becomes harder. This finding suggests that countries may underestimate their vulnerabilities by focusing exclusively on upstream access. Moreover, by tracing how these vulnerabilities evolve over time (2007–2023), we show that systemic trade risk has increased in all clusters, even when exposure and import concentration remained constant. 



One of the most consequential findings of our study is the systematic divergence between dependency indicators for input products and for the full product set, both in the network structure and in their effects on subsequent comparative advantage development. Specifically, input products show higher exposure and supplier concentration than the broader set, yet lower systemic trade risk, while their dependencies affect export growth differently than aggregate dependencies.

Higher exposure and concentration in input products indicate that countries often rely on a small number of dominant suppliers for critical intermediate goods. This reflects the specialized industrial nature of rare-earth-based inputs, where capital intensity, technical know-how, and scale limit diversification. 
Furthermore, our regression results reveal that high baseline exposure across all products is positively associated with a country's subsequent RCA growth, implying import substitution and domestic capability development. However, this pattern does not hold for input product exposure, which shows no significant effect. In other words, exposure to difficult-to-substitute inputs does not reliably translate into export strength absent the requisite industrial base.

In contrast, high supplier concentration in input products correlates positively with RCA growth, whereas concentration across all products does not consistently support export development. This could suggest that concentrated input dependencies may spur industrial policy initiatives, particularly in countries with existing technical capacity, to build domestic midstream capabilities.
However, it might also hint at a lack of diversification and multisourcing efforts in input products as industries develop new export strengths.

Critically, systemic trade risk in input products is significantly negatively associated with RCA changes while at the same time also showing overall lower scores compared to broader set indicators. This underscores that structural vulnerabilities at critical intermediate tiers create persistent barriers to technological upgrading, even when countries seem diversified at the trade level.
These critical vulnerabilities highlight a dimension of risk that indicators for direct dependencies (concentration, exposure, etc.) that do not take value chain aspects into account, fail to capture.

A decrease in the political stability of countries can also result in an increase in systemic trade risk. Although there is a global trend towards decreasing political stability, the changes observed in our study are much smaller than the increases in systemic trade risk. Furthermore, China's political stability indicator has remained relatively stable with some fluctuations, suggesting that changes in political stability alone are unlikely to be the driving factor behind our results.

When aligned with cluster characteristics, these dynamics reveal clear strategic positions in the global rare earth landscape. Cluster 0 (China) exhibits low exposure, concentration, and systemic risk in both input and aggregate profiles, while exerting significant influence, demonstrating China's gatekeeper role in the critical midstream stages. Clusters 1 and 2 (resource-scarce or developing economies) face extreme input exposure and systemic risk, with little capacity for mitigation. Here, dependency does not translate into industrial agency, but rather locks them into vulnerability. Cluster 3 shows decreasing input concentration and rising capability, suggesting potential for selective upgrading. Targeted policies might support these countries in building input capabilities to achieve broader competitiveness. Cluster 4 (advanced economies such as EU, USA, Taiwan, Korea) demonstrates moderate dependency but persistent input-level systemic risk. These countries often have export strengths yet remain exposed to chokepoints in intermediate goods.

At the same time, several limitations need to be acknowledged. HS6 product codes may not capture the full complexity of product transformations, and the LLM-based embedding matches may introduce noise or semantic drift. Future work could integrate firm-level data or material flow models to improve resolution. Moreover, causal inference designs (e.g., natural experiments following trade restrictions) could better establish how dependencies shape comparative advantage over time.

Our framework offers a replicable method for other critical raw materials by integrating AI-assisted code mapping, trade statistics, and network theory. While we focus on rare earths, the same approach can be extended to lithium, cobalt, or gallium—materials with similar geopolitical and industrial relevance. Furthermore, the tier-based risk analysis can help identify where policy interventions (e.g., recycling, diversification, trade agreements) will be most effective. For instance, targeting risk reduction at tier 2 may yield higher resilience gains than investing in upstream extraction alone.

Collectively, the divergent results for input vs all-product dependency metrics, both in network structure and regression outcomes, reveal that strategic dependencies are tier-specific, and not all dependency leads to opportunity. Exposure at the raw material or aggregate level may be compensated, but high systemic risk and lack of processing capacity at the input tier can block technological sovereignty and export development.

\section{Conclusion}\label{sec13}

We advance the understanding of rare earth dependencies by introducing a multilayer, AI-augmented trade-production network framework that captures indirect vulnerabilities across value chain tiers. We reveal that strategic dependencies often reside in intermediate input products, not just upstream raw materials. Further, different dependency indicators behave fundamentally differently when measured at input versus aggregate levels. Finally, systemic trade risk in input tiers constrains export competitiveness, even among advanced economies. 

Our findings suggest that industrial and trade policy must evolve beyond securing raw material access to target the resilience and capacity of key intermediate production nodes. This integrated, tier-aware perspective offers a robust foundation for building more resilient, sovereign, and technologically advanced supply chains in the rare-earth domain and beyond. 

\backmatter


\bmhead{Acknowledgements}

On behalf of the Supply Chain Intelligence Institute Austria (ASCII), we acknowledge financial support from the Austrian Federal Ministry for Economy, Energy and Tourism (BMWET) and the Federal State of Upper Austria. PK and MH acknowledge financial support from the FFG KIRAS project DAGMAR.

\bibliography{sn-bibliography}

\end{document}